\documentclass[a4paper]{article}

\usepackage{sectsty} 
\usepackage{amssymb}        
\usepackage{amsmath}        
\usepackage{graphicx}
\usepackage{epsfig}        
\usepackage{psfrag}
\usepackage{amsthm}

\subsectionfont{\small}

\theoremstyle{definition}\newtheorem{definition}{Definition}
\theoremstyle{plain} 
\theoremstyle{plain}

\title{Observable effects in a class of spherically symmetric static Finsler spacetimes}
\author{Claus L{\"a}mmerzahl\thanks{ZARM, University of Bremen, 28359 Bremen, 
Germany. Email: claus.laemmerzahl@zarm.uni-bremen.de} { , } 
Volker Perlick\thanks{ZARM, University of Bremen, 28359 Bremen, 
Germany. Email: volker.perlick@zarm.uni-bremen.de} { and }
Wolfgang Hasse\thanks{Institute for Theoretical Physics, TU Berlin, Sekr. EW 7-1, 10623
Berlin, Germany; and Wilhelm Foerster Observatory Berlin, 12169 Berlin, Germany. Email: astrometrie@gmx.de.}
}

\date{}

\begin{document}

\vspace{-1cm}
\maketitle

\vspace{-0.7cm}

\begin{abstract}
After some introductory discussion of the definition of Finsler spacetimes
and their symmetries, we consider a class of spherically symmetric and static
Finsler spacetimes which are small perturbations of the Schwarzschild spacetime. 
The deviations from the Schwarzschild spacetime are encoded in three 
perturbation functions $\phi _0(r)$, $\phi _1(r)$ and $\phi _2(r)$  
which have the following interpretations: $\phi _0$ perturbs the time 
function, $\phi _1$ perturbs the radial length measurement and $\phi _2$ 
introduces a spatial anisotropy which is a genuine Finsler feature.   
We work out the equations of motion for freely falling particles and for light rays, 
i.e. the timelike and lightlike geodesics, in this class of spacetimes, and we 
discuss the bounds placed on the perturbation functions by observations in the 
Solar system.  
\end{abstract}

\maketitle

\section{Introduction}\label{sec:intro}

Since its discovery almost hundred years ago, general relativity has proven to
give  a very succesful description of our universe. Nonetheless, there are good
reasons for investigating gravitational theories that are more general than
general relativity. 
There are many
theoretical predictions, in particular from quantum gravity ideas, that general 
relativity should be replaced by a more general theory at some scale. In order to 
confront such theoretical predictions with experiments, it is necessary to 
theoretically study all observable effects of the more general theory. This will
tell by what sort of future experiments deviations from general relativity could 
be observed, and to what accuracy general relativity is verified by present day
observation. The PPN formalism provides a mathematical framework for doing so;
however, it is restricted to metrical theories in the strict sense, i.e., to theories 
where the gravitational field is described by a pseudo-Riemannian metric tensor of 
Lorentzian signature, as in general relativity. For other theories, no 
such universal framework exists.  

In this paper we want to investigate Finsler gravity theories, i.e., theories where
the pseudo-Riemannian metric of general relativity is replaced with a Finsler 
metric. Finsler metrics are characterised by a Lagrangian function that is 
still homogeneous with respect to the velocities, but not necessarily given by a 
quadratic form. The most important feature that distinguishes a Finsler metric 
from a pseudo-Riemannian metric is in the fact that it breaks spatial isotropy 
even in ``infinitesimally small regions'', i.e., mathematically speaking, on the 
tangent space. It is true that up to now there is no observational indication for 
such an anisotropy. (Note, however, Bogovslovsky's \cite{Bogoslovsky1973}
attempt to explain apparent violations of the GZK limit of cosmic rays as an 
effect of a spatial anisotropy.) At the position of the Earth, deviations from 
isotropy are strongly restricted by experiments of the Michelson-Morley type 
\cite{LaemmerzahlLorekDittus2009}. However, this result is based on the 
assumption that the armlength of the Michelson-Morley interferometer is to 
be determined not with the Finsler metric but with an independent Lorentzian 
background metric. If one assumes, by contrast, that the metric which
determines the length of solid bodies shows the same sort of anisotropy
as the metric that determines the light cones, then a Michelson-Morley-type 
experiment would give a null result.   
 
Finsler manifolds have been considered as possible spacetime 
models by a large number of authors.  A fairly complete list of the 
pre-1985 literature can be found in Asanov's book \cite{Asanov1985}.
There are several quite different motivations for considering Finsler 
manifolds as possible spacetime models. Apart from the aesthetic appeal
Finsler geometry has for many authors, Finsler spacetimes have been
recently suggested as a possible explanation for dark matter
\cite{ChangLi2009}, and they have been used in an attempt for explaining 
the Pioneer anomaly \cite{LiChang2009}. (As it has now become clear that
the latter can be explained as a thermal recoil effect 
\cite{RieversLaemmerzahl2011}, also cf. \cite{FranciscoBertolamiGilParamos2012} and \cite{TuryshevTothKinsellaLeeLokEllis2012}, 
this motivation should be considered as obsolete.) At a more fundamental level, 
it has been shown that Finsler geometry naturally comes up in models 
motivated by quantum gravity ideas \cite{GirelliLiberatiSindoni2007}, in 
particular in Very Special Relativity \cite{GibbonsGomisPope2007} and in 
other theories with violation of Lorentz invariance \cite{Kostelecky2011}.  

We take this as the motivation for investigating, in this article, the 
observational bounds on a spherically symmetric and static Finsler 
perturbation of the standard general relativity model of our Solar 
system. To that end we consider the effect such a Finsler perturbation 
would have on the motion of freely falling particles and light rays. In 
contrast to the above-mentioned Michelson-Morley-type experiments, 
we will not need any assumption on the behaviour of (``rigid'') 
extended bodies under the influence of a Finsler perturbation.

As the class of \emph{all} spherically symmetric and static Finsler spacetimes 
is unmanageable (see Section \ref{sec:sym} below), we need a special ansatz.  
As we want to discuss 
the motion of particles and of light rays, we need  a Finsler spacetime in which 
both timelike and lightlike geodesics are well defined. This is an important 
issue, because in the literature one can find many Finsler spacetimes in
which the notion of lightlike geodesics is \emph{not} well defined. As this 
fact is glossed over in many articles, we discuss it in Section \ref{sec:sym} 
below in some detail. Roughly speaking, three different definitions
of Finsler spacetimes can be found in the literature: The one most 
frequently, though often implicitly,  used in physics texts can be found in
Asanov's book \cite{Asanov1985}; an alternative one is due to Beem 
\cite{Beem1970} and a quite recent one, which is a generalisation of Beem's,
is due to Pfeifer and Wohlfarth \cite{PfeiferWohlfarth2011}. As we will 
outline in Section \ref{sec:sym} below, none of them is appropriate for
our purpose: Asanov's definition does not allow to define lightlike geodesics,
while Beem's definition is slightly too restrictive to define staticity in the most 
convenient way; the latter observation is unaffected by Pfeifer and Wohlfarth's 
generalisation. Therefore, we will introduce our own definition of Finsler
spacetimes in Section \ref{sec:sym} below which is a slight generalisation
of Beem's definition. On the basis of this definition, we will then consider in
Section \ref{sec:class} a special class of Finsler spacetimes that are 
perturbations of the Schwarzschild metric. The perturbations preserve 
spherical symmetry and staticity. In this way we arrive at a formalism 
that allows us to quantitatively study, in Section \ref{sec:circ} to \ref{sec:orbits},
hypothetical Finsler deviations from general relativity in the Solar system, not 
only in terms of effects on particles but also on light rays; as in general relativity, 
our light rays are defined as geodesics whose initial vectors lie on a unique light 
cone that determines the causal structure of spacetime. We believe that such a 
formalism did not exist before. It is true that Roxburgh \cite{Roxburgh1992} 
set up a  PPN formalism for Finsler gravity, cf. Roxburgh and Tavakol 
\cite{RoxburghTavakol1979} for related material. This, however, was restricted 
to the very special case of a Finsler metric whose light cones coincide with the
light cones of a pseudo-Riemannian metric; thereby any Finsler effect on the 
lightlike geodesics was excluded. There is also work by Aringazin
and Asanov \cite{AringazinAsanov1985, Asanov1992} on Finsler generalisations 
of the Schwarzschild metric and possible observable effects. However, this is based 
on Asanov's definition for which lightlike geodesics are not defined. Earlier work by 
Coley \cite{Coley1982} is also (implicitly) based on Asanov's definition. More
recently, Pfeifer and Wohlfarth \cite{PfeiferWohlfarth2012} have considered 
a certain Finsler perturbation of the linearised Schwarzschild metric; however,
their ansatz is quite different from ours insofar as it introduces birefringence.

In analogy to the PPN formalism, our analysis will be purely kinematical,
not using any field equation. Several attempts of establishing a Finsler 
generalisation of Einstein's (vacuum) field equation have been brought forward, 
see Rund and Beare \cite{RundBeare1972} (cf. Asanov  \cite{Asanov1985}, 
pp 110), Rutz \cite{Rutz1993}, and  Pfeifer and Wohlfarth \cite{PfeiferWohlfarth2012}.
However, it seems fair to say that no generally accepted Finsler version of a field 
equation exists so far. 

As an aside, we mention that Finsler spacetimes in the sense considered in this paper
provide a counter-example to the Schiff conjecture. In its original version, brought
forward by L. Schiff in 1960 \cite{Schiff1960}, this conjecture said that a theory 
must satisfy Einstein's equivalence principle if it satisfies the weak equivalence
principle. In our Finsler spacetimes there is a unique timelike geodesic for every 
timelike initial condition, so the weak equivalence principle is satisfied. However,
as the theory is not based on a pseudo-Riemannian metric, Einstein's equivalence 
principle is violated.   

\section{Definition of Finsler spacetimes and their symmetries}\label{sec:sym}

Historically, Finsler geometry was first established for positive definite metrics.
In this case, which is covered in standard text-books such as Rund \cite{Rund1959},
a Finsler structure is defined in terms of a function $F(x,\dot{x})$ that is positive
and sufficiently smooth on the set of all tangent vectors $(x, \dot{x})$ with
$\dot{x} \neq 0$, and positively homogeneous of degree one, i.e., 
$F(x,k \dot{x}) = k F(x, \dot{x})$ for all $k > 0$. 
The Finsler metric is then introduced as the Hessian
\begin{equation}\label{eq:g}
g_{\mu \nu} ( x , \dot{x} ) \, = \, \dfrac{\partial ^2 \mathcal{L} (x ,\dot{x} )}{
\partial \dot{x}{}^{\mu} \partial \dot{x}{}^{\nu}},
\end{equation} 
where $\mathcal{L} ( x , \dot{x} ) = F(x, \dot{x})^2$, and it is required that 
this be positive definite for all $\dot{x} \neq 0$. The affinely parametrised
Finsler geodesics are the solutions to the Euler-Lagrange equations of the
Lagrangian $\mathcal{L} ( x , \dot{x})$,
\begin{equation}\label{eq:Euler}
\frac{d}{dt} 
\Big( \frac{\partial {\mathcal{L}}}{\partial {\dot{x}}{}^{\mu}} \Big)
\, = \, \frac{\partial {\mathcal{L}}}{\partial x ^{\mu}} \; .
\end{equation}
For applications to spacetime theory
one would like to have a Finsler metric of Lorentzian signature, and one would 
like to have timelike, lightlike and spacelike geodesics. This requires a 
modified definition of Finsler structures where the Lagrangian $\mathcal{L}$ 
is no longer positive (hence not the square of a real-valued function $F$) on the 
set of \emph{all} non-zero tangent vectors. Therefore
Asanov \cite{Asanov1985} defines a Finsler structure in terms of a function 
$F(x, \dot{x})$ that has the same properties as in the positive definite
formalism, but is given only on some subset of the tangent bundle which he
calls the ``admissible vectors''. The admissible vectors are to be interpreted
as timelike. Typically, such a Finsler function $F$ involves 
the square root of an expression that becomes negative on part of the tangent 
bundle; vectors where this happens are not admissible. In Asanov's formalism, 
timelike geodesics are well-defined as the solutions to the Euler-Lagrange 
equations of the Lagrangian $\mathcal{L} = F^2$ (or $\mathcal{L} = - F^2$,
depending on the choice of signature) with admissible initial conditions. However, 
lightlike geodesics are not well-defined. What one would like to define as lightlike 
vectors are the ones on the boundary of the set of admissible vectors; there, 
however, the Euler-Lagrange equations break down because of zeros in the 
denominator. This formalism of Asanov, in which the Finsler structure is well-behaved 
only on the timelike vectors, is used in many physics papers on indefinite Finsler 
metrics, usually more implicitly than explicitly. The weakness of this approach is in 
the fact that there is no straightforward 
way of defining light rays in this setting. Asanov suggests a notion of light rays 
(see Chapter 7 in \cite{Asanov1985}) that depends on the choice of an auxiliary 
vector field. As the physical meaning of this auxiliary vector field is obscure, we 
do not think that this definition of light rays is satisfactory, from a physical point  
of view. Therefore, as we want to consider the equation of motion of light rays, 
we find Asanov's definition of Finsler structures inappropriate for the purpose
of this paper.

Fortunately, there is an alternative definition. Finsler metrics of Lorentzian signature 
were considered by Beem \cite{Beem1970} in a way that is free from the above-mentioned
drawbacks. In Beem's formalism there is no analogue of the Finsler function $F$; the
Finsler structure is rather given directly in terms of the Lagrangian $\mathcal{L}(x, \dot{x})$,
which should be sufficiently smooth (Beem requires it to be of 
class $C^4$) and real-valued for all $x$ and all $\dot{x} \neq 0$, it should be 
positively homogeneous of  degree two,
\begin{equation}\label{eq:hom}
\mathcal{L}(x,k \dot{x}) = k^2 \mathcal{L}(x, \dot{x}) \quad 
\text{for} \: \,  \text{all} \: \; k > 0 \, ,
\end{equation}
and the Finsler metric (\ref{eq:g}) should be non-degenerate with Lorentzian signature 
for all $\dot{x} \neq 0$. The non-degeneracy condition guarantees that the Euler-Lagrange 
equations (\ref{eq:Euler}) admit a unique solution to any initial condition $(x(0), \dot{x} (0))$ 
with $\dot{x} (0) \neq 0$. These solutions are the affinely parametrised  Finsler geodesics 
which are well-defined for timelike ($\mathcal{L} ( x , \dot{x} ) < 0$), lightlike 
($\mathcal{L} ( x , \dot{x} ) = 0$) and spacelike ($\mathcal{L} ( x , \dot{x} ) > 0$) 
tangent vectors $\dot{x} \neq 0$. So in Beem's setting light rays can be unambiguously 
defined as lightlike geodesics, just as in standard general relativity. This is the reason 
why we consider, in this paper, a Finsler structure in the sense of Beem. Actually, for
reasons that will become clear soon, we find it necessary to generalise Beem's 
definition a little bit: Our Lagrangian $\mathcal{L}$ will not be smooth (and not even 
$C^2$) at all $(x, \dot{x})$ with $\dot{x} \neq 0$; the second derivative of 
$\mathcal{L}$ will give undetermined expressions on a set of measure zero. However, 
there will still be a unique solution curve (geodesic) through each point $(x, \dot{x})$ 
with $\dot{x} \neq 0$.

Another interesting generalisation of Beem's definition was brought 
forward recently by Pfeifer and Wohlfarth 
\cite{PfeiferWohlfarth2011,PfeiferWohlfarth2012}. The main idea of
their work is to allow for Lagrangians $\mathcal{L}$ that are homogeneous of 
\emph{any} degree. However, as they still assume $\mathcal{L}$ to be 
``smooth'' at all $(x, \dot{x})$ 
with $\dot{x} \neq 0$, their generalisation is of no advantage for 
our purpose, although it
might be fruitful for other applications.

Guided by Beem's definition  \cite{Beem1970}
we define a Finsler spacetime in the following way.

\begin{definition}\label{def:Finsler}
A Finsler spacetime is a 4-dimensional manifold $M$ with a Lagrangian function
$\mathcal{L}$ that satisfies the following properties:
\begin{itemize}
\item[(a)]
$\mathcal{L}$ is a real-valued function on the tangent bundle $TM$ minus the zero section, i.e., 
$\mathcal{L}(x, \dot{x})$ is defined for all $(x, \dot{x})$ with $\dot{x} \neq 0$.
\item[(b)]
$\mathcal{L}$ is positively homogeneous of degree two with respect to $\dot{x}$, i.e.,
eq. (\ref{eq:hom}) holds.
\item[(c)]
The Finsler metric (\ref{eq:g}) is well-defined and has Lorentzian signature $(-+++)$ for almost all 
$(x, \dot{x})$ with $\dot{x} \neq 0$. (As usual, ``almost all'' means ``up to a set of measure zero''.)
\item[(d)]
The Euler-Lagrange equations (\ref{eq:Euler})
admit a unique solution for every initial condition $(x, \dot{x})$ with $\dot{x} \neq 0$;
at points where the Finsler metric is not well-defined this solution is to be constructed
by continuous extension.
\end{itemize}
\end{definition} 
On a Finsler spacetime, we represent points in $M$ by their coordinates $x = (x^0,x^1,x^2,x^3)$ 
and points in the fibre $T_xM$ of the tangent bundle by their induced coordinates $\dot{x} = 
(\dot{x}{}^0,\dot{x}{}^1,\dot{x}{}^2,\dot{x}{}^3)$. We use Einstein's summation convention 
for greek indices taking values 0,1,2,3 and for latin indices taking values 1,2,3. 

Note that the homogeneity condition (b) of the Lagrangian implies that the Finsler metric is
positively homogeneous of degree zero,
\begin{equation}\label{eq:homg}
g_{\mu \nu} ( x , k \, \dot{x} ) \, = \, g_{\mu \nu} ( x , \dot{x} ) \quad \text{for \; all} \: k > 0 \; ,
\end{equation}
and that the Lagrangian can be written in terms of the Finsler metric as
\begin{equation}\label{eq:Lg}
\mathcal{L}(x, \dot{x}) \, = \, \dfrac{1}{2} \, g_{\mu \nu} ( x , \dot{x} ) 
\dot{x}{}^{\mu} \dot{x}{}^{\nu} \; .
\end{equation}
With the help of the Lagrangian we classify non-zero tangent vectors as timelike 
($\mathcal{L}(x, \dot{x})<0$), lightlike ($\mathcal{L}(x, \dot{x})=0$) or spacelike 
($\mathcal{L}(x, \dot{x})>0$). We call the solutions to the Euler-Lagrange 
equations (\ref{eq:Euler}) the affinely parametrised Finsler \emph{geodesics}. Again
by the homogeneity condition (b) of the Lagrangian, $\mathcal{L}(x, \dot{x})$ is a 
constant of motion; hence Finsler geodesics can be classified as timelike, lightlike or 
spacelike. We interpret the timelike geodesics as freely falling particles and the lightlike 
geodesics as light rays. This interpretation is in agreement with the idea that the
(Finsler) spacetime geometry tells freely falling particles and light rays how to move, i.e.,
that no additional mathematical structures enter into the equations of motion for freely
falling particles and light rays. We have already mentioned that some authors disagree 
with this hypothesis, as far as light rays are concerned. As the interpretation of lightlike
Finsler geodesics as light rays is crucial for our work, some additional justification is
given in the Appendix.

In this paper we want to consider a special class of Finsler spacetimes that will serve us as a 
model for the gravitational field around the Sun. We shall assume that this gravitational field
is static and spherically symmetric. In order to make these notions precise we have to recall
that symmetries of Finsler metrics are described in terms of (Finsler generalisations of)
Killing vector fields. By definition, a vector field $V = V^{\mu} \partial / \partial x^{\mu}$ on 
a Finsler spacetime $M$ is a \emph{Killing vector field} if and only if its flow, if lifted to $TM$, 
leaves the Lagrangian $\mathcal{L}$ invariant. This condition can be rewritten in terms of the 
Finsler metric as
\begin{equation}\label{eq:Killing}
V^{\mu} \dfrac{\partial g_{\rho \sigma}}{\partial x^{\mu}} +
\dfrac{\partial V^{\tau}}{\partial x^{\nu}} \dot{x}{}^{\nu} 
 \dfrac{\partial g_{\rho \sigma}}{\partial \dot{x}{}^{\tau}} +
\dfrac{\partial V^{\tau}}{\partial x^{\rho}} g_{\tau \sigma}
+
\dfrac{\partial V^{\tau}}{\partial x^{\sigma}} g_{\rho \tau} \; = \; 0 \; .
\end{equation}
Here the $V^{\mu}$ depend on $x$ only, whereas the $g_{\mu \nu}$ depend on $x$ and 
$\dot{x}$. The Finslerian Killing equation (\ref{eq:Killing}) is known since the early days 
of Finsler geometry, see Knebelman \cite{Knebelman1929}. 

In the standard formalism of general
relativity, one defines a spacetime as \emph{stationary} if it admits a timelike Killing
vector field $V$ and as \emph{static} if, in addition, this timelike Killing vector field $V$ is
orthogonal to hypersurfaces. If we exclude global pathologies (such as, e.g., the case
that the quotient space $M/V$ fails to be a Hausdorff manifold), the latter condition
implies that the spacetime is a warped product of a 3-dimensional manifold with a
(positive definite) Riemannian metric and the real line with a negative definite 
metric. We can use this property as the definition of staticity for Finsler spacetimes.

\begin{definition}\label{def:static} 
A Finsler spacetime $(M ,\mathcal{L} )$ is \emph{static} if $M$ is diffeomorphic to a 
product, $M \simeq \mathbb{R} \times N$, and $\mathcal{L}$ is of the form
\begin{equation}\label{eq:static}
\mathcal{L}(x^1,x^2,x^3,\dot{t}, \dot{x}{}^1 ,  \dot{x}{}^2 ,  \dot{x}{}^3 )
\, = \, \dfrac{1}{2} \, \Big( \, g_{tt}(x^1,x^2,x^3) \, \dot{t}{}^2 \, + \, 
g_{ij}(x^1,x^2,x^3, \dot{x}{}^1 ,  \dot{x}{}^2 ,  \dot{x}{}^3)  \, \dot{x}{}^i
 \dot{x}{}^j \, \Big) \, ,
\end{equation}
where $t$ runs over $\mathbb{R}$ and $(x^1,x^2,x^3)$ are coordinates on $N$;
the temporal metric coefficient $g_{tt}(x^1,x^2,x^3)$ must be negative and the
spatial metric $g_{ij}(x^1,x^2,x^3, \dot{x}{}^1 ,  \dot{x}{}^2 ,  \dot{x}{}^3)$
must be positive definite.
\end{definition}

If the $g_{ij}$ are independent of the $\dot{x}{}^i$, Definition \ref{def:static}
reduces to the definition of a static spacetime in the sense of general relativity.
In any other case the limit of the $g_{ij}$ for $( \dot{x}{}^1, \dot{x}{}^2,
\dot{x}{}^3) \to (0,0,0)$ depends on the direction in which this limit is 
performed; this follows immediately from eq. (\ref{eq:homg}). As a consequence, 
the Finsler metric fails to be well-defined on vectors tangent to the $t$-lines. 
This is the reason why, in part (c) of Definition \ref{def:Finsler}, the restriction
to ``almost all'' non-zero tangent vectors was necessary to include proper
Finsler Lagrangians of the form of eq. (\ref{eq:static}).

We now add the condition of spherical symmetry. By definition, a Finsler spacetime is 
spherically symmetric if it admits a 3-dimensional algebra of Killing vector fields that
generate the rotation group $\mathrm{SO}(3)$ such that each of its orbits is 
diffeomorphic to the 2-sphere $\mathrm{S}^2$. For a static Finsler spacetime as
given in eq. (\ref{eq:static}), spherical symmetry means that we can choose the
spatial coordinates as $x^1=r,x^2=\vartheta , x^3= \varphi$, where $r$ labels 
the group orbits and $\vartheta$ and $\varphi$ are standard coordinates on $S^2$,
and that then $g_{tt}$ depends on $r$ only and the spatial part $g_{ij}\dot{x}{}^i
\dot{x}{}^j$ depends on $r$, $\dot{r}$ and $\dot{\vartheta}{}^2
+ \mathrm{sin}^2 \vartheta \, \dot{\varphi}{}^2$ only. For a derivation of 
the latter fact see McCarthy and Rutz \cite{McCarthyRutz1993,Rutz1996}.

\section{A class of spherically symmetric and static Finsler spacetimes}\label{sec:class}

As the class of all spherically symmetric and static Finsler spacetimes is too big, we
make a more special ansatz for our model of the Solar system. We assume 
that the Lagrangian $\mathcal{L}$ is of the form
\begin{equation}\label{eq:defL}
2 \, {\mathcal{L}}=  
\big( 
h_{tt} + c^2 \psi _0 \big) 
{\dot{t}}{}^2 + \Big( \big( h_{ij} h_{kl} +
\psi _{ijkl} \big) {\dot{x}}{}^i {\dot{x}}{}^j {\dot{x}}{}^k {\dot{x}}{}^l
\Big)^{\frac{1}{2}} \; .
\end{equation}
Here 
\begin{equation}\label{eq:defh}
h_{tt} dt^2 + h_{ij} dx^i dx^j = h_{tt} dt^2 +  h_{rr} dr^2
+  r^2 \big( {\mathrm{sin}}^2 \vartheta \,
d \varphi ^2 + d \vartheta ^2 \big)
\end{equation}
is a spherically symmetric and static Lorentzian metric. In this section and in
the following one, $h_{tt}$ and $h_{rr}$ are arbitrary functions of $r$, 
but later they will be specified to be the Schwarzschild metric coefficients,
\begin{equation}\label{eq:schwarz}
h_{tt}  \, = \, - \, \dfrac{c^2}{h_{rr}} \, = \,
 - \, c^2 \, \big( 1 - \frac{\, 2 \, G \, M \,}{c^2 \, r} \big)
\; , 
\end{equation}
where $c$ is the speed of light, $G$ is the gravitational constant, and $M$ is
the mass of the gravitating body. The spatial perturbation $\psi _{ijkl}$ is 
spherically symmetric and independent of $t$,
\begin{equation}\label{eq:psi}
\psi _{ijkl} {\dot{x}}{}^i {\dot{x}}{}^j {\dot{x}}{}^k {\dot{x}}{}^l =
\psi _1(r) {\dot{r}}{}^4 + \psi _2(r) r^2 {\dot{r}}{}^2 \big( {\mathrm{sin}}^2 \vartheta \, 
{\dot{\varphi}}{} ^2 + {\dot{\vartheta}}{}^2 \big)
+ \psi _3(r) r^4 \big( {\mathrm{sin}}^2 \vartheta \,
{\dot{\varphi}}{} ^2 + {\dot{\vartheta}}{} ^2 \big)^2
\end{equation}
and the time perturbation $\psi _0$ is a function of $r$ only.

Actually, ansatz (\ref{eq:defL}) is less special than it might appear. 
The fourth-order term $\psi _{ijkl}  {\dot{x}}{}^i {\dot{x}}{}^j {\dot{x}}{}^k 
{\dot{x}}{}^l$ can be viewed as the leading order term in 
a general Finsler power--law perturbation of the spatial  part of the metric.
(We do not want to consider a third-order term because it would violate the 
symmetry under spatial inversions $\dot{x}{}^i \to - \dot{x}{}^i$.)
For this reason, we consider the Lagrangian 
(\ref{eq:defL}) as a natural choice for our purpose. 

In the following we refer to the dimensionless quantities $\psi _A(r)$ as 
to the ``perturbation functions'', $A=0,1,2,3$. Throughout this paper, we assume that the 
perturbation functions depend differentiably on $r$ and are so small that we 
may linearise all equations with respect to the $\psi _A(r)$ and their derivatives 
$\psi '_A(r)$. Differentiability and smallness of the $\psi _A(r)$ guarantee that
the Lagrangian ${\mathcal{L}}(x , \dot{x})$ is real-valued, and the Finsler metric 
(\ref{eq:g}) is non-degenerate with Lorentzian signature for \emph{almost} all 
$(x, {\dot{x}})$ with ${\dot{x}} \neq 0$ . The only points where
this condition is violated are the points where the \emph{spatial} velocity
components are all zero, $(\dot{x}{}^1, \dot{x}{}^2 , \dot{x}{}^3) = (0,0,0)$,
but $\dot{t} \neq 0$. At these points, the Finsler metric gives undetermined
expressions. We will see in the next section that, even through these points, the 
solutions to the Euler-Lagrange equations are uniquely determined by continuous
extension, i.e., that our ansatz gives indeed a Finsler spacetime in the sense of 
Definition \ref{def:Finsler}. 

We are still free to transform the radial coordinate. We can remove this freedom,
thereby reducing the number of perturbation functions from four to three. In the
unperturbed (Schwarzschild) spacetime, $r$ is an ``area coordinate'', i.e., the
area of the sphere at $r$ is given by $4 \pi r^2$. We can fix the radial coordinate 
by requiring that $r$ has the same geometric meaning in the perturbed spacetime.
From equations (\ref{eq:defL}) and (\ref{eq:psi}) we read that, in the perturbed
spacetime, the sphere at $r$ has area $4 \pi r^2 (1+ \psi _3)$. Hence, the desired 
condition is satisfied if we allow only perturbations with $\psi _3 =0$. We are then
left with three perturbation functions  $\psi _0$, $\psi _1$ and $\psi _2$, and the
Lagrangian (\ref{eq:defL}) reads 
\begin{equation}\label{eq:defL2}
2 \, {\mathcal{L}}=  
\big( h_{tt} + c^2 \psi _0 \big) {\dot{t}}{}^2 + 
\Big(  \sqrt{h_{rr}^{\; 2} + \psi _1} \dot{r}{}^2+r^2 
\big( {\mathrm{sin}}^2 \vartheta \, 
{\dot{\varphi}}{} ^2 + {\dot{\vartheta}}{}^2 \big) \Big)
\sqrt{ \, 1 + \, \dfrac{\big( 2 h_{rr} - 2 \sqrt{h_{rr}^{\; 2}+ \psi _1} + \psi _2 \big)
r^2 \dot{r}{}^2 \big( {\mathrm{sin}}^2 \vartheta \, 
{\dot{\varphi}}{} ^2 + {\dot{\vartheta}}{}^2 \big)
}{
\Big( \sqrt{h_{rr}^{\; 2} + \psi _1} \dot{r}{}^2+r^2 
\big( {\mathrm{sin}}^2 \vartheta \, 
{\dot{\varphi}}{} ^2 + {\dot{\vartheta}}{}^2 \big) \Big) ^2
} }  \; .
\end{equation}
From this expression we read that $\mathcal{L}$ is the Lagrangian of a 
pseudo-Riemannian metric if and only if
\begin{equation}\label{eq:Riemann}
2 h_{rr} - 2 \sqrt{h_{rr}^{\; 2} + \psi _1 } + \psi _2 \, = \, 0  \, .
\end{equation}
In this case the equations of motion can be investigated in terms of the 
standard PPN formalism. If (\ref{eq:Riemann}) does not hold, we have a 
proper Finsler geometry and the PPN formalism does not apply. We might
say that the left-hand side of eq. (\ref{eq:Riemann}) measures the
``Finslerity'' of our perturbed spacetime.

\section{Equations of motion}

We now discuss the solutions to the Euler-Lagrange equations (\ref{eq:Euler}) 
in our class of spherically symmetric and static Finsler spacetimes, i.e., the 
affinely parametrised Finsler geodesics. We restrict to timelike (${\mathcal{L}} < 0$) 
and lightlike (${\mathcal{L}} = 0$) geodesics, which are to be interpreted as freely 
falling particles and as light rays, respectively. For timelike geodesics we can fix the 
parametrisation by requiring $2{\mathcal{L}} = -c^2$; then the affine parameter 
is equal to Finsler proper time $\tau$.

By symmetry, it suffices to consider particles and 
light rays in the equatorial plane $\vartheta = \pi /2$. Then the linearised
version of the Lagrangian (\ref{eq:defL2}) reads
\begin{equation}\label{eq:linL}
2 {\mathcal{L}}= (1+ \phi _0)  
h_{tt} {\dot{t}}{}^2 + 
(1+\phi_1) h_{rr} {\dot{r}}{}^2
+ r^2 {\dot{\varphi}}{}^2 +
\dfrac{\phi _2 h_{rr} r^2 {\dot{r}}{}^2  {\dot{\varphi}}{}^2
}{
h_{rr}  {\dot{r}}{}^2 + r^2 {\dot{\varphi}}{}^2 } \; .
\end{equation}
Here we have introduced, for notational convenience, modified perturbation
functions
\begin{equation}\label{eq:phi}
\phi _0 = \dfrac{c^2 \psi _0}{h_{tt}} \; , \quad
\phi _1 = \dfrac{\psi _1}{2 h_{rr}^{\; 2}} \; , \quad
\phi _2 = \dfrac{\psi _2 h_{rr} - \psi _1}{2 h_{rr}^{\; 2}} \; .
\end{equation}
In terms of these modified perturbation functions, and after linearisation,
the ``non-Finsler condition'' (\ref{eq:Riemann}) simply reads $\phi _2 =0$. 
Hence, in the linearised setting the ``Finslerity`` of our perturbed
spacetime is measured just by $\phi _2$.

By equation (\ref{eq:linL}), each of the perturbation functions $\phi _0$, $\phi _1$ 
and $\phi _2$ has an obvious interpretation: $\phi _0$ perturbs the time function $t$,
$\phi _1$ perturbs the radial length measurement and $\phi _2$ introduces a
spatial anisotropy which is a genuine Finsler feature.   Circular 
motion ($\dot{r} =0$) feels only $\phi _0$ while radial motion 
($\dot{\varphi} = 0$) feels $\phi _0$ and $\phi _1$; the ``Finslerity''
$\phi _2$ is felt only by motion that is neither circular nor radial.  

Equation (\ref{eq:linL}) is the form of the Lagrangian on which all our 
following results are based. We will now derive the equations of motion.
 
In addition to the constant of motion
\begin{equation}\label{eq:Lconst1}
{\mathcal{L}}= - \frac{c^2}{2} \: \text{ for freely falling particles }
\end{equation}
or
\begin{equation}\label{eq:Lconst2}
{\mathcal{L}}= 0 \: \text{ for light, }
\end{equation}
the $t$ and $\varphi$ components of the Euler-Lagrange equations
give two more constants of motion $E$ and $L$,
\begin{equation}\label{eq:energy}
- \, E \, = \, \frac{\partial {\mathcal{L}}}{\partial {\dot{t}}} 
\, = \, (1+\phi_0) \, h_{tt} \, {\dot{t}} \; ,
\end{equation}
\begin{equation}\label{eq:angularm}
L \, = \, \frac{\partial {\mathcal{L}}}{\partial {\dot{\varphi}}} 
\, = \, r^2 \, \dot{\varphi} \,
\left( \, 1 \, + \, \dfrac{ \phi _2 \, h_{rr}^{\; 2} \dot{r}{}^4
}{
\big( h_{rr}  {\dot{r}}{}^2 + r^2 \, {\dot{\varphi}}{}^2 \big)^2} \, 
\right) 
\; .
\end{equation}
The three constants of motion $\mathcal{L}$, $E$ and $L$ give us 
three equations that determine
the geodesics. From these three equations we read that, by continuity,
there is a unique geodesic even for initial conditions $\dot{r} (0) = 0$, 
$\dot{\varphi} (0) =0$ and $\dot{t} (0) \neq 0$, for which the 
Euler-Lagrange equations yield undetermined expressions, namely 
a curve with $\varphi = \mathrm{constant}$. This completes the proof 
that our Lagrangian defines a Finsler spacetime in the sense of 
Definition \ref{def:Finsler}.

To within our linear approximation, the three conservation equations 
(\ref{eq:linL}), (\ref{eq:energy}) and (\ref{eq:angularm}) can be solved 
for $\dot{t}$, $\dot{\varphi}$ and $\dot{r}{}^2$, 
\begin{equation}\label{eq:dott}
\dot{t} \, = \, \dfrac{- E}{h_{tt}} \, \big( \, 1 \, - \, \phi _0 \, \big) \; ,
\end{equation}
\begin{equation}\label{eq:dotphi}
\dot{\varphi} \, = \, \dfrac{L}{r^2} \, \left(\, 1 \, - \,
\dfrac{ \phi_2 \,
\Big( 2 \mathcal{L} - \dfrac{E^2}{h_{tt}}-\dfrac{L^2}{r^2} \Big) ^2
}{
\Big( 2 \mathcal{L} - \dfrac{E^2}{h_{tt}} \Big)^2} \, \right) \; ,
\end{equation}
\begin{equation}\label{eq:dotr}
\dot{r}{}^2 \, = \, \dfrac{1}{h_{rr}} \, 
\Big( 2 \mathcal{L} - \dfrac{E^2}{h_{tt}}-\dfrac{L^2}{r^2} \Big) 
\left( \, 1 \, - \, \phi _1 \, + \, \dfrac{\phi_2 \, L^2
\Big( 2 \mathcal{L} - \dfrac{E^2}{h_{tt}}-\dfrac{2 \, L^2}{r^2} \Big) 
}{
r^2 \Big( 2 \mathcal{L} - \dfrac{E^2}{h_{tt}} \Big) ^2} \right) 
\, + \, \dfrac{\phi_0 E^2}{h_{tt} h_{rr}} \; .
\end{equation}
From these three equations we find
\begin{equation}\label{eq:eomt2}
\frac{d \varphi}{dt} \, = \, \dfrac{\dot{\varphi}}{\dot{t}} 
\, = \, 
\dfrac{\,L \, h_{tt}}{- E \, r^2} 
\, \left( \, 1 \, + \, \phi _0 \, - \, 
\dfrac{
\phi _2 \, \Big( 2 \mathcal{L} - \dfrac{E^2}{h_{tt}}-\dfrac{L^2}{r^2} \Big) ^2
}{
\Big( 2 \mathcal{L} - \dfrac{E^2}{h_{tt}} \Big) ^2}
\, \right) \; ,
\end{equation}
\begin{equation}\label{eq:eomt3}
\big( \dfrac{dr}{dt} \big)^2 \, = \, 
\dfrac{{\dot{r}}{}^2}{{\dot{t}}{}^2} \, = \,
\dfrac{
h_{tt}^{\; 2}}{E^2 \, h_{rr}} \, 
\Big( 2 {\mathcal{L}} - \dfrac{E^2}{h_{tt}}- \dfrac{L^2}{r^2} \, \Big)
\left(
\, 1 \, + \, \phi _0 \, - \, \phi _1 \, + \, 
\dfrac{\, \phi _2 \, L^2
\Big( 2 {\mathcal{L}} - \dfrac{E^2}{h_{tt}}- \dfrac{2L^2}{r^2} \, \Big)
}{
r^2 \Big( 2 {\mathcal{L}} - \dfrac{E^2}{h_{tt}} \Big) ^2} \right)
\, + \, \dfrac{\phi _0 h_{tt}^{\; 2} \Big( 2 {\mathcal{L}} - \dfrac{L^2}{r^2} \, \Big)
}{E^2 h_{rr}} \; .
\end{equation}
Equations (\ref{eq:eomt2}) and (\ref{eq:eomt3}) determine the trajectories if parametrised 
by coordinate time $t$. 
If we are interested only in the geometrical shape of the trajectory, but not in 
its parametrisation, we may use the equation
\begin{equation}\label{eq:eomphi}
( \dfrac{dr}{d \varphi} )^2 \, = \, \dfrac{\dot{r}{}^2}{\dot{\varphi}{}^2} \, = \, 
\dfrac{r^4 \, 
\Big( 2 \, {\mathcal{L}} - \dfrac{E^2}{h_{tt}}- \dfrac{L^2}{r^2} \Big)}{
h_{rr} \, L^2} 
\left( \, 1 - \, \phi _1 \, +
\dfrac{\, \phi _2 
\Big( 4 \, {\mathcal{L}} - \dfrac{2 \, E^2}{h_{tt}}- \dfrac{3 \, L^2}{r^2} \, \Big)
}{
\Big( 2 \, {\mathcal{L}} - \dfrac{E^2}{h_{tt}} \Big) } \right)
\, + \, \dfrac{\phi _0 \, E^2 r^4}{
h_{rr} h_{tt} L^2} \; .
\end{equation}

\section{Circular orbits}\label{sec:circ}
For a particle ($2 \mathcal{L} = - c^2$) on a circular orbit, the equations
$dr/d \varphi = 0$ and $d^2r/d \varphi ^2=0$ must hold.  By equation
(\ref{eq:eomphi}), these two conditions are equivalent to
\begin{equation}\label{eq:circle1}
-c^2-\dfrac{E^2}{h_{tt}}- \dfrac{L^2}{r^2} + 
\dfrac{\phi _0 E^2}{h_{tt}} \, = \, 0 \, ,
\end{equation}
\begin{equation}\label{eq:circle2}
\dfrac{E^2}{h_{tt}^{\; 2}} \, h_{tt} ' \, + \, 
\dfrac{2 \, L^2}{r^3} \, + \, E^2 \, \Big( \dfrac{\phi_0'}{h_{tt}} \, - \, 
\dfrac{\phi _0 \, h_{tt}'}{h_{tt}^{\; 2}} \, \Big)  \, = \, 0 \, .
\end{equation}
With $E^2$ and $L^2$ determined this way, equation (\ref{eq:eomt2})
yields
\begin{equation}\label{eq:circle3}
\Big( \dfrac{d \varphi}{dt} \Big) ^2 \, = \, 
- \, \dfrac{h'_{tt}}{2 \, r} \, \Big\{ \, 1 \, +  \, 
\dfrac{\big( \phi _0 h_{tt} \big) '}{h_{tt}'} \, \Big\} \; .
\end{equation}
After inserting the Schwarzschild metric (\ref{eq:schwarz}), 
we find
\begin{equation}\label{eq:circle4}
\Big( \dfrac{d \varphi}{dt} \Big) ^2 \, = \, 
\dfrac{GM}{r^3} \, \Big\{ \, 1 \, + \,   
\dfrac{c^2r^2}{2GM}  \left( \phi _0 \Big( 1 - \dfrac{2GM}{c^2 r} \Big) \right) ' 
\, \Big\} \; .
\end{equation}
If we denote the period by $T$, we have $d\varphi /dt = 2 \pi /T$, and 
(\ref{eq:circle4}) gives a generalisation of the third Kepler law for
circular orbits,
\begin{equation}\label{eq:kepler}
\frac{r^3}{T^2} \, 
\Big\{ \, 1 \, - \, 
\dfrac{c^2r^2}{2GM}  \left( \phi _0 \Big( 1 - \dfrac{2GM}{c^2 r} \Big) \right) '
\, \Big\}
\, = \, \frac{\, G \, M \,}{\, 4 \, \pi ^2  \, } \; .
\end{equation} 
In the unperturbed Schwarzschild spacetime, the Kepler law
\begin{equation}\label{eq:kepler0}
\frac{r^3}{T^2} \,  = \, \frac{G \, M}{\, 4 \, \pi ^2 \,} 
\end{equation} 
coincides with the Newtonian Kepler law, as is well known.

From an experimentalist's point of view, one may use the unperturbed Kepler
law (\ref{eq:kepler0}) as an operational definition of $GM$. 
According to standard general relativity this would lead to a constant 
value $GM$. With our perturbation, it would lead to an $r$-dependent
value $\widehat{GM} (r)$ that is related to the constant $GM$ value
from general relativity by  
\begin{equation}\label{eq:GF}
\widehat{GM}  (r) \,  = \, GM \, 
\Big\{ \, 1 \, + \, 
\dfrac{c^2r^2}{2GM}  \left( \phi _0 \Big( 1 - \dfrac{2GM}{c^2 r} \Big) \right) '
\, \Big\} \; .
\end{equation} 
We will now discuss the bounds imposed on $\phi _0 (r)$ by (\ref{eq:GF}).

From observations, the 
gravitational constant $G$ is known, at present, only up to a relative 
uncertainty of approximatively $10^{-4}$. However, our knowledge of 
the product $GM$, where $M$ denotes the Solar mass, is much better. 
The most recent value, taken from the webpage of the Jet Propulsion 
Laboratory  {\tt http://ssd.jpl.nasa.gov/?constants}, is
\begin{equation}\label{eq:GM}
GM = 1.32712440018 \times 10^{20} \mathrm{m}^3 \mathrm{s}^{-2} 
\, \pm \, 
8 \times 10^9 \, \mathrm{m}^3 \mathrm{s}^{-2} \; .
\end{equation}
More specifically, we get a value for $GM= 4 \pi ^2 a^3/T^2$ 
from the observed values of the semi-major axis $a$ and of the 
period $T$ for each individual planet.  As the periods are known with
a higher accuracy than the semi-major axes we can write
\begin{equation}\label{eq:epsilon}
\Big| \, \dfrac{\widehat{GM} (a) \, - \, GM}{GM} \, \Big|
\, \lessapprox \,  3 \, \epsilon  
\end{equation}
where $\epsilon$ is the accuracy with which the semi-major axis $a$
is known. The values of $\epsilon$ for the eight planets are shown in
Table \ref{table:epsilon}.

\begin{table}[h]
\begin{center}
\begin{tabular}{|c|c|c|c|c|c|c|c|c|}
\hline
$\begin{matrix} \, \\ \, \end{matrix}$ &  Mercury  &  Venus  & Earth  
& Mars  &  Jupiter  &  Saturn  &  Uranus  &  Neptune \\[0.2cm]
\hline
$a/(10^{10}\mathrm{m})$ & $\begin{matrix} \, \\ \, \end{matrix} 5.79 
\begin{matrix} \, \\ \, \end{matrix}$ & $10.8 $ & 
$15.0 $ & $22.8$ &
$77.9 $ & $143 $ & 
$288$ & $450 $ 
\\[0.2cm]
\hline
$\Delta a/\mathrm{m}$ & $\begin{matrix} \, \\ \, \end{matrix} 0.11  
\begin{matrix} \, \\ \, \end{matrix}$ & $0.33 $ & $0.15 $ & 
$0.66 $ & $640$ & $4200$ & 
$3.8 \! \times \! 10^4$ & $4.8 \! \times \! 10^5$ 
\\[0.2cm]
\hline
$\epsilon$ & $\! \begin{matrix} \, \\ \, \end{matrix} 1.9 \! \times \! 10^{-11}  
\begin{matrix} \, \\ \, \end{matrix} \! $ & $\! 3.1 \! \times \!10^{-11} \! $ & 
$\! 1.0 \! \times \! 10^{-11} \! $ & $\! 2.9 \! \times \! 10^{-11} \! $ & 
$\! 8.2 \! \times \! 10^{-9} \! $ & $\! 2.9 \! \times \! 10^{-8} \! $ & 
$\! 1.3 \! \times \! 10^{-7} \! $ & $\! 1.1 \! \times \! 10^{-6} \!$ 
\\[0.2cm]
\hline
\end{tabular}
\end{center}
\caption{The first row of this table gives the semi-major axis $a$ of each planet and
the second row gives the formal standard deviation $\Delta a$ of $a$. We have taken 
the values for $\Delta a$ from Table 4 in Pitjeva \cite{Pitjeva2005} who reported on
data determined by the Institute of Applied Astronomy of the Russian Academy of
Sciences from observations made between 1913 and 2003. In the third row we 
calculated the accuracy $\epsilon$ of the semi-major axis for each planet, assuming
that the real error may be one order of magnitude bigger than the formal standard
deviation, $\epsilon = 10 \Delta a/a$.
\label{table:epsilon}
}
\end{table}

In Section \ref{sec:orbits} below we will consider non-circular orbits. 
We will see that then $\phi _1$ and $\phi _2$ do have an effect. For 
the time being we will be satisfied with a rough order-of-magnitude 
estimate given by replacing each of the planets with a hypothetical 
planet that moves on a circular orbit with $r=a$. Then we can compare 
(\ref{eq:GF}) with (\ref{eq:epsilon}) and 
conclude that
\begin{equation}\label{eq:f0bounds}
\left|   \left( \phi _0 \Big( 1 - \dfrac{2GM}{c^2 r} \Big) \right) ' \right|
\, \lessapprox \, 
\dfrac{2 \, GM}{c^2r^2} \, 3 \epsilon \; .
\end{equation}
Integration from $r_1$ to $r_2$ yields
\begin{equation}\label{eq:f0bounds2}
\left|   \left( \phi _0 (r_2) \Big( 1 - \dfrac{2 \, GM}{c^2 r_2} \Big) -
\phi _0 (r_1) \Big( 1 - \dfrac{2 \,GM}{c^2 r_1} \Big)\right)  \right|
\, \lessapprox \, 
\dfrac{2 \, GM \, \big| r_2-r_1 \big|}{c^2 \, r_1^2 \, r_2^2} \, 
3 \, \epsilon _{\mathrm{max}}
\end{equation}
where $\epsilon _{\mathrm{max}}$ is the maximal uncertainty between 
$r_1$ and $r_2$. 
As $2 \, GM/(c^2r)$ varies from $10^{-8}$ near the Mercury orbit
to $10^{-10}$ near the Neptune orbit, (\ref{eq:f0bounds2}) implies
\begin{equation}\label{eq:f0bounds3}
r_1 \, \left|   
\dfrac{\phi _0 (r_2) - \phi _0 (r_1)}{r_2 \, - \, r_1}
 \right|
\, \lessapprox \, 
10^{-16}
\end{equation}
for all radii $r_1$ and $r_2$ between the Mercury orbit and the Neptune 
orbit. If both $r_1$ and $r_2$ are between the Mercury orbit and the Mars
orbit, the bound is even three orders of magnitude smaller. In particular, 
\begin{equation}\label{eq:f0bounds4}
r  \, \left| \phi _0 ' (r)  \right|
\, \lessapprox \, 
10^{-16}
\end{equation}
between Mercury and Neptune, and even three orders of magnitude smaller
between Mercury and Mars. Note that no assumption on monotonicity of the 
function $\phi _0$ was needed for this result.

In analogy to the PPN formalism, one could assume that the perturbation
functions are of the form
\begin{equation}\label{eq:falloff}
\phi _A (r) \, = \, \phi _{A1} \, \dfrac{2GM}{c^2r} \, + \,
O \Big( \big( \dfrac{2GM}{c^2r} \big) ^2 \Big) \, , \qquad A=0,1,2
\end{equation}
with constants $\phi _{A1}$. We have included the Schwarzschild
radius $2GM/c^2$ in this ansatz to make the $\phi _{A1}$ dimensionless.
Here we mean by $GM$ the constant of Nature given by the fixed 
numerical value after the equality sign in (\ref{eq:GM}), just as we
mean by $c$ the constant of Nature given by the numerical value
$299 \, 792 \, 458$ m/s. If we accept the ansatz (\ref{eq:falloff}) and
if we assume that, as a reasonable first approximation, the terms of 
second and higher order can be neglected, the inequality 
(\ref{eq:f0bounds}) yields
\begin{equation}\label{eq:phi01a}
\big| \phi _{01} \big| \, \Big( \, 1 \, - \, \dfrac{4GM}{c^2 r} \, \Big)
\, \lessapprox \, 3 \, \epsilon \; .
\end{equation}
With the best value of $\epsilon$ taken from Table \ref{table:epsilon}
we find   
\begin{equation}\label{eq:phi01b}
\big| \phi _{01} \big|
\, \lessapprox \, 3 \times 10^{-11} \; .
\end{equation}

\section{Radial free fall}\label{sec:ff}
In this section we consider a freely falling particle ($2 {\mathcal{L}} = - c^2$) 
that moves in the radial direction, i.e. ${\dot{\varphi}}=0$. By (\ref{eq:angularm}), 
the latter condition is equivalent to $L=0$, so (\ref{eq:eomt3})
simplifies to
\begin{equation}\label{eq:radial1}
\Big( \dfrac{dr}{dt} \Big) ^2 \, = \, \dfrac{h_{tt}^{\; 2}}{h_{rr} E^2} \,
\Big( - c^2 - \, \dfrac{E^2}{h_{tt}} \, \Big) \, 
\big( 1+ \phi _0 -\phi _1 \big) \, - \, 
\dfrac{\phi _0 c^2 h_{tt}^{\; 2}}{h_{rr} E^2}  \; .
\end{equation}
This expression gives the particle's velocity as it is measured by static observers 
with clocks that show coordinate time $t$. If we want to consider the same observers 
with clocks that show (Finsler) proper time $\tau$, we have to use the relation
\begin{equation}\label{eq:proper}
c^2 \, d\tau ^2 \, = \, - \, (1+ \phi _0 ) \,  h_{tt} \, dt^2 \; . 
\end{equation}
Then (\ref{eq:radial1}) can be rewritten as
\begin{equation}\label{eq:radial2}
\Big( \dfrac{dr}{d \tau} \Big) ^2 \, = \, \dfrac{c^2 h_{tt}}{E^2 h_{rr}} \, \Big\{
\Big( c^2 + \, \dfrac{E^2}{h_{tt}} \, \Big) \,  \,
\big( 1-\phi _1 \big) \, + \, c^2 \, \phi _0 \,  \Big\} \; .
\end{equation}
As a first application of this equation we want to discuss the acceleration 
of a particle from rest. From (\ref{eq:radial2}) we read that, at a point where
$dr/d \tau=0$, the equation 
\begin{equation}\label{eq:rest}
c^2 \, + \, \dfrac{E^2}{h_{tt}}  \, + \, c^2 \, \phi _0  \, = \, 0 
\end{equation}
must hold. By differentiating (\ref{eq:radial2}) with respect to $\tau$, and inserting
(\ref{eq:rest}) afterwards, we find the following expression for the acceleration from
rest.
\begin{equation}\label{eq:radial3}
2 \,  \dfrac{d^2r}{d \tau ^2} \, = \, - \, \dfrac{c^2 h_{tt}'}{h_{tt} h_{rr}} \, 
\Big\{
\, 1 \, - \, \phi _1  \, + \, \phi _0 ' \, \dfrac{h_{tt}}{h_{tt}'} \,  \Big\} \; .
\end{equation}
With the Schwarzschild metric (\ref{eq:schwarz}) this can be rewritten as
\begin{equation}\label{eq:radial4}
\dfrac{d^2r}{d \tau ^2} \, = \, - \, \dfrac{GM}{r^2} \, 
\Big\{ \, 1 \, - \, \phi _1  \, - \, \phi _0 ' \, r \, + \, 
\phi _0 ' \dfrac{c^2r^2}{2GM} \,  \Big\} \; .
\end{equation}
In the unperturbed Schwarzschild spacetime (\ref{eq:radial4}) reduces to
\begin{equation}\label{eq:acc5}
\frac{d^2 r}{d \tau ^2} \, = \, 
- \, \frac{\, G \, M \,}{r^2}  \; .
\end{equation}
As an alternative to the method discussed in Section
\ref{sec:circ}, an experimentalist could use (\ref{eq:acc5}) as
an operational definition of $GM$. According to standard general
relativity, this would lead to the same constant value $GM$ as
the method of Section \ref{sec:circ}. With our 
perturbation, however, it would lead to an $r$-dependent value
\begin{equation}\label{eq:GF2}
\widetilde{GM} (r)  \,  = \, GM \, 
\Big\{ \, 1 \, - \, \phi _1  (r) \, - \, \phi _0 ' (r) \, r \, + \, 
\phi _0 ' (r) \dfrac{c^2r^2}{2GM} \,  \Big\}
\end{equation} 
which is different from the $\widehat{GM} (r)$ of (\ref{eq:GF}),
\begin{equation}\label{eq:GF3}
\widehat{GM} (r)  \, - \, \widetilde{GM} (r)  \,  = \, GM \, 
\big\{ \, \phi _0 (r) \, + \, \phi _1 (r) \,  \big\} \; .
\end{equation} 
We see that the perturbation functions $\phi _0$ and $\phi _1$ 
can be determined, in principle, by observing circular orbits
and radial acceleration from rest. With the bounds on $\phi _0$ 
we have found from the observation of circular orbits in 
Section \ref{sec:circ}, we can now discuss the bounds on 
$\phi _1$ that result from measurements of free-fall accelerations.
Again, we are satisfied with a rough order-of-magnitude estimate.
We can then say that, with the Pioneer anomaly explained as
a thermal recoil effect, all observations of radial accelerations 
up to the Neptune orbit are in agreement with General Relativity
to within the order of the anomalous Pioneer acceleration of 
$9 \times 10^{-10} \mathrm{m/s}{}^2$. At the Neptune orbit, $GM
/r^2 \approx 6 \times 10^{-3} \mathrm{m/s}{}^2$. As a consequence,
(\ref{eq:GF2}) suggests that   
\begin{equation}\label{eq:GF4}
\Big| \, \phi _1 (r) \, - \, \dfrac{c^2r}{GM} \, r \, \phi _0 ' (r) \,
\Big( \, 1 \, - \, \dfrac{2 \, GM}{c^2r} \Big)  \Big|
 \, \lessapprox \, \dfrac{9 \times 10 ^{-10}}{6 \times 10^{-3}} \; .
\end{equation} 
Using $|a|-|b| \le \big| |a|-|b| \big| \le |a-b|$,  we find 
\begin{equation}\label{eq:GF4a}
\big| \phi _1 (r) \big| \, \lessapprox \, 
\Big| \, \dfrac{c^2r}{GM} \, r \, \phi _0 ' (r) \,
\Big( \, 1 \, - \, \dfrac{2 \, GM}{c^2r} \Big)  \Big|
 \, + \, 1.5 \times 10^{-7} \; .
\end{equation} 
From Section \ref{sec:circ} we know that $\Big|
\dfrac{c^2r}{2 \, GM} \, r \, \phi _0 ' (r) \Big| \lessapprox 10^{-6}$ 
between the Mercury orbit and the Neptune orbit, hence
\begin{equation}\label{eq:GF5}
\big| \phi _1 (r) \big| \,  \lessapprox \, 10 ^{-6} 
\end{equation} 
for all $r$ in this range. 

If we assume that the perturbation functions have  a fall-off
behaviour according to (\ref{eq:falloff}), and if we neglect
terms of second and higher order, (\ref{eq:GF4a}) implies
\begin{equation}\label{eq:phi11a}
\big| \phi _{11} \big| \, \dfrac{2GM}{c^2r} \, \lessapprox \,  
2 \,  \big| \phi_{01} \big| \, \Big( \, 1 \, - \, \dfrac{2 \, GM}{c^2r} \Big)  
\, + \, 1.5 \times 10^{-7} \; .
\end{equation}
Evaluating at the Mercury orbit, $ \dfrac{2GM}{c^2 r} \approx 5 \times 10^{-8}$,
and using (\ref{eq:phi01b}) yields
\begin{equation}\label{eq:phi11b}
 \big|  \phi _{11} \big|  \, \lessapprox \, 3 \; .
\end{equation}
This is much less restrictive than the bound on $\phi _{01}$ we had found 
before. Note, however, that a value of $\phi _{11}$ of the order of unity
gives only a small correction to $h_{rr}$ of the Schwarzschild metric, because 
our ansatz (\ref{eq:falloff}) involves the Schwarzschild radius.
  
\section{Effects on the paths of light rays}\label{sec:light}

In this section we want to calculate the effect of our Finsler 
perturbation on the worldline of a light ray that comes in 
from a source at radial coordinate $r_S$, passes the Sun at a 
minimal value $r_m$ of the radial coordinate, and goes out 
again to an observer at radial coordinate $r_O$.  

To that end we have to evaluate (\ref{eq:eomt3}) and 
(\ref{eq:eomphi}) with the Schwarzschild metric 
coefficients (\ref{eq:schwarz}) and $\mathcal{L} =0$.
For notational convenience, we will use the abbreviation
\begin{equation}\label{eq:defp}
p(r) \, = \, r^{-2} \, 
\Big( 
\, 1 \, - \, \dfrac{2GM}{c^2r}
\, \Big) 
\end{equation}
throughout. With this abbreviation, and $\mathcal{L} = 0$,
equation (\ref{eq:eomphi}) can be rewritten in the following
form.
\begin{equation}\label{eq:lightphi}
\Big( \dfrac{dr}{d \varphi} \Big) ^2 \, = \,
r^4 \, \Big( \dfrac{E^2}{c^2L^2} - p \Big) 
\Big\{ 1 - \phi _1 + 2 \, \phi _2  - 3 \, \phi _2 c^2 p  \dfrac{L^2}{E^2}
\Big\} \, - \, \dfrac{\phi _0 r^4 E^2}{c^2L^2} \; .
\end{equation}
We first determine how the constants of motion $E$ and 
$L$ depend on the minimum radius $r_m$. If we insert the 
value $r=r_m$ into the right-hand side of (\ref{eq:lightphi}), 
we must get zero. If we solve the resulting equation for
$E^2/L^2$, we find
\begin{equation}\label{eq:lEL}
\dfrac{E^2}{L^2} \, = \, c^2 p(r_m) 
\Big\{ 1 + \phi_0 (r_m)  \Big\} \; .
\end{equation}
Inserting this value into (\ref{eq:lightphi}) yields
\begin{equation}\label{eq:lphi}
\big( \frac{dr}{d \varphi} \big)^2 \, = \, A(r)
\big( \, 1 - \, \alpha (r) \, \big)
\end{equation}
where
\begin{equation}\label{eq:defA}
A(r) \, = \, 
r^4 \, \big( \, p(r_m) \, \, - \, p(r) \, \big) \,
\end{equation}
and
\begin{equation}\label{eq:defalpha}
\alpha (r) \, = \, 
 \phi _1 (r) 
-\phi _2 (r) \dfrac{2 p (r_m) - 3 p(r) }{p(r_m)} -
\big( \phi _0 (r_m) - \phi _0 (r) \big) 
\dfrac{p(r_m)}{p(r_m ) - p(r)} \; .
\end{equation}

An analogous calculation puts (\ref{eq:eomt3}) into the form
\begin{equation}\label{eq:lt}
\big( \frac{dr}{dt} \big)^2 \, = \, B(r)
\big( \, 1 - \, \beta (r) \, \big)
\end{equation}
where
\begin{equation}\label{eq:defB}
B(r) \, = \,  
c^2 \, r^4 \, p(r)^2 \, 
\Big( \, 1 \, - \, \dfrac{p(r)}{p(r_m)} \, \Big) \, 
\end{equation}
and
\begin{equation}\label{eq:defbeta}
\beta (r) \, = \, 
 \phi _1 (r) -\phi _2 (r) p(r) \dfrac{p (r_m) - 2 p(r) }{p(r_m)^2} -
\phi _0 (r)  -  \big( \phi _0 (r_m)  -\phi _0 (r)  \big) 
\dfrac{ p(r)}{p(r_m)-p(r)} \; .
\end{equation}
Note that the perturbations $\alpha (r)$ and $\beta (r)$
depend not only on $\phi _0(r)$, $\phi _1(r)$ and $\phi _2(r)$ but also
on $\phi _0(r_m)$, because of (\ref{eq:lEL}).

\subsection{Light deflection}\label{subsec:deflect}

From (\ref{eq:lphi}) we find
\begin{equation}\label{eq:ld2}
d \varphi \, = \, 
\Big( \, 1 \, + \, \dfrac{\alpha (r)}{2}  \, \Big) \; 
\dfrac{
dr
}{
\, \sqrt{A(r)} \,
}
\end{equation}
and integration yields the deflection angle $\Delta \varphi$,
\begin{equation}\label{eq:dangle1}
\pi \, + \, \Delta \varphi \, = \, 
\Big( \, \int _{r_m} ^{r_S} \, + \, \int _{r_m} ^{r_O} \, \Big) \,
\Big( \, 1 \, + \, \frac{\alpha (r)}{2} \, \Big) \, \frac{dr}{\sqrt{A(r)}} \, .
\end{equation}
If we denote by $\Delta \varphi _0$ the deflection angle in the unperturbed 
Schwarzschild spacetime,
\begin{equation}\label{eq:defphi0}
\pi \, + \, \Delta \varphi _0 \, = \, 
\Big( \, \int _{r_m} ^{r_S} \, + \, \int _{r_m} ^{r_O} \, \Big) \,
\frac{dr}{\sqrt{A(r)}} \, ,
\end{equation}
the deflection angle in the perturbed spacetime reads
\begin{equation}\label{eq:phiphi0}
\Delta \varphi \, = \, \Delta \varphi _0 \, + \,  
\Big( \, \int _{r_m} ^{r_S} \, + \, \int _{r_m} ^{r_O} \, \Big) \,
\frac{\alpha (r) \, dr}{2 \, \sqrt{A(r)}} \, .
\end{equation}
Using the mean value theorem, this result can be rewritten as
\begin{equation}\label{eq:dangle2}
\Delta \varphi \, = \, \Delta \varphi _0 \, + \, 
\frac{\alpha ( \tilde{r} )}{2}  \, 
\Big( \, \pi \, + \, \Delta \varphi _0 \, \Big) \, ,
\end{equation}
where $\tilde{r}$ is some radius value between $r_m$ and 
$\mathrm{max}(r_S,r_O)$.

We want to evaluate these equations for the case that the source is a 
distant star ($r_S \to \infty$), the observer is on the Earth ($r_O=1 \,
\mathrm{AU}$), and the light ray is grazing the surface of the Sun ($r_m =
0.0046 \, \mathrm{AU}$). Then (\ref{eq:defphi0}) gives the well-known deflection 
angle of 
\begin{equation}\label{eq:dphi0}
\Delta \varphi _0 \, = \, 1.75'' \, = \, 
8.48 \, \times\, 10^{-6} \, \text{rad} \, 
\end{equation}
and present day observations (see Will \cite{Will2006}, Section 3.4)
require
\begin{equation}\label{eq:angleacc}
\Delta \varphi \, - \, \Delta \varphi _0 \, = \,
 \frac{1}{2} \, \Big( \, - \, 1.7 \, \pm \, 4.5 \, \Big) \times 
10^{-4} \, \Delta \varphi _0 \, .
\end{equation}
Comparison of (\ref{eq:dangle2}) and (\ref{eq:angleacc}) gives a bound
for the possible values of $\alpha (\tilde{r})$,
\begin{equation}\label{eq:alphabounds}
| \alpha ( \tilde{r} ) | \, = \, 
2 \, \Big| \, \frac{\, \Delta \varphi \, - \, 
\Delta \varphi _0 \,}{ \Delta \varphi _0} \, \Big| \;
\Big| \, \frac{\Delta \varphi _0}{ \, \pi \, + \, \Delta \varphi _0} \, \Big|
\, \lessapprox \, 2 \times 10^{-9} \; .
\end{equation}
If we assume that the perturbation functions $\phi _A (r)$ have a 
fall-off behaviour according to eq. (\ref{eq:falloff}), and if we
neglect terms of second and higher order, the integral in (\ref{eq:phiphi0})
can be calculated numerically. For $r_S \to \infty$, $r_O = 1 \, \mathrm{AU}$
and $r_m = 0.0046 \, \mathrm{AU}$ we find
\begin{equation}\label{eq:danglephiA1}
\Delta \varphi \, - \, \Delta \varphi _0 \, = \, 
4.2 \times 10^{-6} \phi _{11} \, + \, 
5.1 \times 10^{-11} \phi _{21} \, - \, 
4.2 \times 10^{-6} \phi _{01} \, .
\end{equation}
We see that $\phi _{21}$ contributes with a much smaller
factor than the other two perturbations. This has its reason in 
the fact that in (\ref{eq:defalpha}) the Finslerity $\phi _2 (r)$
comes with a factor that changes sign between $r=r_m$ and 
$r=r_O$, therefore positive and negative contributions to the
integral partly cancel out. Hence, light deflection 
is rather insensitive to the Finslerity of our spacetime model.
Combining (\ref{eq:danglephiA1}) with (\ref{eq:angleacc}), and 
using (\ref{eq:dphi0}) and (\ref{eq:phi01b}), gives the quite
insignificant bound
\begin{equation}\label{eq:phi11phi21a}
\big| \, 8.2 \times 10^4 \,  \phi _{11} \, + \, 
\phi _{21} \, \big| \, \lessapprox \, 52 \, .
\end{equation}

\subsection{Time delay of light rays}\label{subsec:ltravel}
From (\ref{eq:lt}) we find
\begin{equation}\label{eq:lt2}
d t \, = \, 
\big( \, 1 \, + \, \dfrac{1}{2} \, \beta (r) \, \big) \; 
\dfrac{
dr
}{
\, \sqrt{B(r)} \,
} \; .
\end{equation}
Integration of this equation yields the travel time. The
difference to the Newtonian travel time $t_N$ is, by 
definition, the time delay $\delta t$,
\begin{equation}\label{eq:deltat}
t_N \, + \, \delta t \, = \, 
\Big( \, \int _{r_m} ^{r_S} \, + \, \int _{r_m} ^{r_O} \, \Big)
\big( \, 1 \, + \, \frac{1}{2} \, \beta (r) \, \big) \, 
\frac{dr}{\sqrt{B(r)}} \, .
\end{equation}

In the unperturbed Schwarzschild spacetime, the time delay 
$\delta t_0$ is given by 
\begin{equation}\label{eq:deltat0}
t_N \, + \, \delta t _0 \, = \, 
\Big( \, \int _{r_m} ^{r_S} \, + \, \int _{r_m} ^{r_O} \, \Big)
\frac{dr}{\sqrt{B(r)}} \, ,
\end{equation}
hence
\begin{equation}\label{eq:deltatt0}
\delta t \, = \, \delta t_0 \, + \,  
\Big( \, \int _{r_m} ^{r_S} \, + \, \int _{r_m} ^{r_O} \, \Big)
\frac{ \, \beta (r) \, dr}{2 \, \sqrt{B(r)}} \, .
\end{equation}
Using the mean value theorem, this can be rewritten 
as
\begin{equation}\label{eq:deltatt0m}
\delta t \, = \, \delta t _0 \, + \, \dfrac{1}{2} \, 
\beta (\hat{r}) \big( \, t_N \, + \, \delta t_0 \, \big)
\end{equation}
where $\hat{r}$ is some radius value between $r_m$ and $\mathrm{max}(r_S,r_O)$.

Time delays have been measured with radar signals since the 
1960s. In the beginning Mars, Mercury and Venus were used as
passive reflectors. In this case the round-trip travel time 
for a signal from the Earth to the planet and back is two 
times the one-way travel time (\ref{eq:deltat}) plus a 
correction taking the orbital motion of the Earth into account. 
Later time delay experiments used spacecraft. The most accurate 
experiment of this kind was made with radio signals sent 
to the Cassini spacecraft, with the result (see Will \cite{Will2006}, 
Sec. 3.4) that 
\begin{equation}\label{eq:td}
\dfrac{\, \delta t - \delta t_0 \,}{\delta t_0} \, = \,
(2.1 \pm 2.3) \times 10^{-5} \; .
\end{equation}
The measurement was made when Cassini was at a distance of 8.43 AU from the 
Sun, and the distance of closest approach $r_m$ was 1.6 Solar radii ($=0.0074
\, \mathrm{AU}$). This corresponds to $\delta t_0 \approx 273 \, \mu \mathrm{s}$
and $t_N \approx 4700 \, \mathrm{s}$. Hence we find, with (\ref{eq:deltatt0m}) and 
(\ref{eq:td}), a very small bound for a certain linear combination of the perturbation
functions at some radius value $\hat{r}$ between 0.0074 AU and 8.43 AU,
\begin{equation}\label{eq:betabounds}
| \beta (\hat{r} ) | \, = \, 2 \, \Big| \dfrac{\, \delta t \, - \, \delta t _0 \,}{\delta t_0} \Big|
\, \Big| \dfrac{\delta t_0}{\, t_N\, + \, \delta t_0 \, } \Big| \, \lessapprox \, 
5.2 \times 10^{-12} \; .
\end{equation}
If only the leading-order terms in (\ref{eq:falloff}) are taken into account, numerical
calculation of the integral in (\ref{eq:deltatt0}) yields (for $r_m = 0.0074
\, \mathrm{AU}$, $r_S=8.43 \, \mathrm{AU}$ and $r_O=1 \, \mathrm{AU}$)
\begin{equation}\label{eq:tdphiA1}
\delta t \, - \, \delta t_0  \, = \,
\big( \, 6.6 \times 10^{-5} \, \phi _{11} \, + \, 
3.3 \times 10^{-6} \, \phi _{21} \, - \, 
7.5 \times 10^{-5} \, \phi _{01} \, \big) \, \mathrm{s} \, .
\end{equation}
The left-hand side can be estimated with the help of (\ref{eq:td}). With $\delta t_0
\approx 273 \, \mu \mathrm{s}$ and (\ref{eq:phi01b}) we find
\begin{equation}\label{eq:phi11phi21b}
\big| \, 20 \, \phi _{11} \, + \, \phi _{21} \, \big| \, \lessapprox  \, 
3.6 \times 10^{-3} \, .
\end{equation}

\section{Effects on bound orbits}\label{sec:orbits}

In Section \ref{sec:circ} we have seen that circular orbits are affected 
only by the coefficient $\phi _0$, but not by $ \phi _1$ and $\phi _2$.
In this section we consider non-circular bound orbits, and we will 
investigate how the perturbation functions influence Kepler's third
law and the perihelion precession.

We consider a massive particle ($2 \mathcal{L}=-c^2$) on a bound orbit,
with minimum radius $r_1$ (perihelion) and  maximum radius $r_2$ 
(aphelion). We need to calculate how the constants of motion $E$ and $L$ 
depend on $r_1$ and $r_2$. To that end, we rewrite (\ref{eq:eomphi}) 
for the Schwarzschild metric coefficients, with $2 \mathcal{L}=-c^2$
and using again the abbreviation (\ref{eq:defp}), in the following form.
\begin{equation}\label{eq:eomphip}
\Big( \dfrac{dr}{d \varphi} \Big) ^2 \, = \, 
\dfrac{r^4 \, \Big( E^2 - c^4r^2p-L^2c^2p \Big)}{c^2 \, L^2}
\left( \, 1 \, - \, \phi_1 \, + \, 2 \, \phi _2 \, - \, 
\dfrac{3 \, \phi_2 L^2 c^2 p}{(E^2-c^4r^2p)} \right) 
\, - \, \dfrac{\phi _0 E^2 r^4}{c^2L^2}
\; .
\end{equation}
For $r=r_1$ and $r=r_2$ the right-hand side of (\ref{eq:eomphip}) has to vanish,
\begin{equation}\label{eq:r1r2}
E^2 - c^4r_1^2p(r_1)-L^2c^2p(r_1) \, + \, \phi _0 (r_1)E^2 \, = \, 0 \, ,
\qquad
E^2 - c^4r_2^2p(r_2)-L^2c^2p(r_2) \, + \, \phi _0 (r_2)E^2 \, = \, 0 \, .
\end{equation}
Solving for $E^2$ and $L^2$ yields
\begin{equation}\label{eq:E2}
E^2 \, = \, E_0^2 \, + \, E_0^2 \, 
\left( 
\, \dfrac{\phi _0(r_2)p(r_1)- \phi _0(r_1)p(r_2)}{p(r_1) \, - \, p(r_2)} \,
\right) 
\; ,
\end{equation}
\begin{equation}\label{eq:L2}
L^2 \, = \, L_0^2 \, + \, 
\dfrac{E_0^2}{c^2} \, \left(
\dfrac{\phi _0(r_2)-\phi _0(r_1)}{p(r_1)-p(r_2)}  \right) 
 \; ,
\end{equation}
where
\begin{equation}\label{eq:EL02}
E_0^2 \, = \, 
\dfrac{
c^4 p(r_1)p(r_2)(r_2^2-r_1^2)}{p(r_1)-p(r_2)}
\qquad \mathrm{and} \qquad
L_0^2 \, = \, 
\dfrac{
c^2 \big( p(r_2)r_2^2-p(r_1)r_1^2 \big)}{
p(r_1)-p(r_2)}
\end{equation}
are the values of the constants of motion in the unperturbed Schwarzschild
spacetime. Substitution of (\ref{eq:E2}) and (\ref{eq:L2}) into 
(\ref{eq:eomphip}), and using the identities $E_0^2 = L_0^2c^2p(r_1) 
+c^4p(r_1)r_1^2 = L_0^2c^2p(r_2) + c^4p(r_2)r_2^2$,
gives the orbit equation in terms of $E_0^2$ and $L_0^2$, 
\begin{equation}\label{eq:eomphipEL}
\Big( \dfrac{dr}{d \varphi} \Big) ^2 \, = \, 
\dfrac{r^4 \, \Big( E_0^2 - c^4r^2p-L_0^2c^2p \Big)}{c^2 \, L_0^2}
\left( \, 1 \, - \, \phi_1 \, + \, 2 \, \phi _2 \, - \, 
\dfrac{3 \, \phi_2 L_0^2 c^2 p}{(E_0^2-c^4r^2p)} 
\right.
\end{equation} 
\[
\left. - \, 
\dfrac{c^2E_0^2 \Big\{ 
\big( \phi _0 (r_2) - \phi _0 (r) \big)
\big( p(r_1)r_1^2-p(r)r^2 \big) 
- \big( \phi _0 (r_1) - \phi _0 (r) \big)
\big( p(r_2)r_2^2-p(r)r^2 \big) \Big\} 
}{
L_0^2 \big( p(r_1)-p(r_2) \big) 
\Big( E_0^2 - c^4r^2p-L_0^2c^2p \Big)}
\right) 
\; .
\]
After substituting $E_0^2$ and $L_0^2$ from  
(\ref{eq:EL02}) we find
\begin{equation}\label{eq:ophi}
\big( \frac{dr}{d \varphi} \big)^2 \, = \, 
C(r) \, 
\big( \, 1 - \, \gamma(r) \big)
\end{equation}
where
\begin{equation}\label{eq:defC}
C(r) \, = \,
\dfrac{ r^2 ( r_2 - r) (r-r_1)}{r_1 \, r_2} \,  
\left( \, 1 \, - \, \dfrac{2GM}{c^2} 
\Big( \, \dfrac{1}{r_1} \, + \, \dfrac{1}{r_2} \, + \, \dfrac{1}{r} \, \Big) \, \right)
\end{equation}
and
\begin{equation}\label{eq:defgamma}
\gamma(r) \, = \, 
\phi _1 (r) \, - \, 2 \, \phi_2 (r)  \, + \, 
\dfrac{
3 \, \phi _2 (r) r_1r_2 \Big( 1- \dfrac{2GM}{c^2r}\Big)
}{
r(r_1+r_2-r) \left( 1 - \dfrac{2GM
\big(r_1^2+r_2^2+r_1r_2-rr_1-rr_2 \big)
}{c^2r_1r_2 \big( r_1+r_2-r \big)} \right)
}
\end{equation}

\[
+ \, 
\dfrac{ 
c^2r(r_1+r_2) \Big( 1- \dfrac{2GM}{c^2r_1} \Big) \Big(1- \dfrac{2GM}{c^2r_2} \Big)
}{
2GM (r_2-r_1) 
\Big\{ 1 - \dfrac{2GM}{c^2} \Big( \dfrac{1}{r_1}+ \dfrac{1}{r_2} + \dfrac{1}{r} \Big) \Big\}
}
\left(
\dfrac{r_2 \big( \phi _0 (r_2) - \phi _0 (r) \big)}{r_2-r} 
-
\dfrac{r_1 \big( \phi _0 (r_1) - \phi _0 (r) \big)}{r_1-r} 
\right)
\; .
\]
For the limiting case of a circular orbit, $r=r_1=r_2$, we find
\begin{equation}\label{eq:circgamma}
\gamma(r) \, = \, 
\phi _1 (r) \, +  \, \phi_2 (r)  \, + \, 
\dfrac{ 
c^2r^2 \Big( 1- \dfrac{2GM}{c^2r} \Big) ^2
}{
GM  
\Big( 1 - \dfrac{6GM}{c^2r}  \Big)
}
\Big( \, \phi_0 ' (r) \, + \,\dfrac{r}{2} \, \phi _0 '' (r) \, \Big)
\; .
\end{equation}
Analogously, substitution of (\ref{eq:E2}) and (\ref{eq:L2}) into 
(\ref{eq:eomt3}) results in 
\begin{equation}\label{eq:ot}
\big( \frac{dr}{d t} \big)^2 \, = \, 
D(r) \, 
\big( \, 1 - \, \delta (r) \big)
\end{equation}
 where
\begin{equation}\label{eq:defD}
D(r) \, = \,
\dfrac{ 2 \, G \, M \, ( r_2 - r) (r-r_1) \, 
\left( \, 1 \, - \, \dfrac{2GM}{c^2r} \, \right)^2 \,  
\left( \, 1 \, - \, \dfrac{2GM}{c^2} 
\Big( \, \dfrac{1}{r_1} \, + \, \dfrac{1}{r_2} \, + \, \dfrac{1}{r} \, \Big) \, \right)
}{
r ^2 \, (r_1+r_2) \, 
\left( \, 1 \, - \, \dfrac{2GM}{c^2r_1} \, \right)
\, 
\left( \, 1 \, - \, \dfrac{2GM}{c^2r_2} \, \right)}
\end{equation}

and

\begin{gather}\label{eq:defdelta}
\delta(r) \, = \, 
\gamma (r) \, - \, 2 \, \phi _0 (r) \, - \, 
\dfrac{c^2 r_1 r_2 \big( \phi _0 (r_2) - \phi _0 (r_1) \big)}{2GM ( r_2-r_1)}
\, + \,
\dfrac{r_2 \phi _0(r_2) - r_1 \phi _0 (r_1)}{r_2-r_1}
\\[0.2cm]
\nonumber
+ \, 
\dfrac{2 \, \phi _2 (r) \, (r_2-r)^2 (r-r_1)^2 
\left( \, 1 \, - \, \dfrac{2GM}{c^2} 
\Big( \, \dfrac{1}{r_1} \, + \, \dfrac{1}{r_2} \, + \, \dfrac{1}{r} \, \Big) \, \right)^2}{
r^2 \left( \, r_1 \, + \, r_2 \, - \, r \, - \, \dfrac{2GM}{c^2 r_1 r_2} 
\big(  r_1^2 + r_2^2 + r_1r_2  - r(r_1+r_2) \big) \, \right)^2} \; .
\end{gather}
For the limiting case of a circular orbit, $r=r_1=r_2$, we find
\begin{equation}\label{eq:circdelta}
\delta (r) \, - \, \gamma (r) \, = \, - \phi _0 (r) \, - \, \dfrac{c^2r^2}{2GM} \,
\Big( \, 1 \, - \, \dfrac{2GM}{c^2r} \, \Big) \, \phi _0 ' (r) \; .
\end{equation} 
Equations (\ref{eq:circgamma}) and (\ref{eq:circdelta}) can be used as valid 
approximations for orbits whose eccentricity is not too big, where $r$ is any
value between the perihelion and the aphelion. 

\subsection{Perihelion precession}\label{subsec:peri}

From (\ref{eq:ophi}) and (\ref{eq:ot}) we find  
\begin{equation}\label{eq:ophi2}
d \varphi \, = \, 
\Big( \, 1  + \, \dfrac{1}{2} \, \gamma (r) \, \Big)
\, \frac{dr}{\sqrt{C(r)}} \; ,
\end{equation}
\begin{equation}\label{eq:ot2}
d t \, = \, 
\Big( \, 1  + \, \dfrac{1}{2} \, \delta (r) \, \Big)
\, \frac{dr}{\sqrt{D(r)}} \; .
\end{equation}
Integrating these two equations over the orbit from one
perihelion transit to the next, 
\begin{equation}\label{eq:defPhi}
2 \, \pi \, + \, \Delta \Phi \, = \, 
2 \, \int _{r_1} ^{r_2} \, 
\Big( \, 1  + \, \dfrac{1}{2} \, \gamma (r) \, \Big) 
\, \frac{dr}{\sqrt{C(r)}} \, ,
\end{equation}
\begin{equation}\label{eq:defT}
T \, = \, 
2 \, \int _{r_1} ^{r_2} \, 
\Big( \, 1  + \, \dfrac{1}{2} \, \delta (r) \, \Big) 
\, \frac{dr}{\sqrt{D(r)}} \, ,
\end{equation}
gives the anomalistic period $T$ (in terms of coordinate time $t$)
and the angular advance $\Delta \Phi$ of the perihelion during this
period. We denote the corresponding quantities in the unperturbed
Schwarzschild spacetime by an index $0$, 
\begin{equation}\label{eq:Phi0}
2 \, \pi \, + \, \Delta \Phi _0 \, = \, 
2 \, \int _{r_1} ^{r_2} \, 
\frac{dr}{\sqrt{C(r)}} \, ,
\end{equation}
\begin{equation}\label{eq:T0}
T _0 \, = \, 
2 \, \int _{r_1} ^{r_2} \, 
\frac{dr}{\sqrt{D(r)}} \, .
\end{equation}
The precession rate of the perihelion, in radians per time, is
$\omega = \Delta \Phi /T$. In our linearised setting $\omega$ 
deviates from $\omega _0 = \Delta \Phi _0 / T_0$ according to
\begin{equation}\label{eq:omega}
\dfrac{\omega - \omega _0}{\omega _0} \, = \, 
\dfrac{1}{\Delta \Phi _0} \, \int _{r_1} ^{r_2} 
\dfrac{\gamma (r) \, dr}{\sqrt{C(r)}} \, - \, 
\dfrac{1}{T _0} \, \int _{r_1} ^{r_2} 
\dfrac{\delta (r) \, dr}{\sqrt{D(r)}} \, . 
\end{equation}
With the mean-value theorem the last equation can be 
rewritten as
\begin{equation}\label{eq:omegam}
\dfrac{\omega - \omega _0}{\omega _0} \, = \, 
\dfrac{\gamma ( \hat{r})}{2} \, 
\Big( \, \dfrac{ 2 \pi}{\Delta \Phi _0} \, + \, 1 \, \Big)
\, - \, \dfrac{\delta ( \tilde{r})}{2} \, ,
\end{equation}
where $\hat{r}$ and $\tilde{r}$ are some radius values
between $r_1$ and $r_2$. 

For Mercury, the precession rate $\omega _0$ according to general relativity 
is well-known to be 43 arcseconds per century. This corresponds to a precession 
angle (in radians) per revolution of $\Delta \Phi _0 = 0.502 \times 10^{-6}$.  
Present day observations confirm that the general-relativistic value is true, with 
a possible relative error of $10^{-3}$,  see Will \cite{Will2006}, Section 3.5. 
Hence, (\ref{eq:omegam}) implies
\begin{equation}\label{eq:omegagam}
\big| 6.2 \times 10^6 \, \gamma ( \hat{r}) 
\, - \, 0.5 \, \delta ( \tilde{r}) \, \big| \, \le \, 10 ^{-3} \, ,
\end{equation}
which is a restrictive bound on $\gamma (\hat{r})$ for $\hat{r}$ on the Mercury 
orbit.

If we take only leading-order terms of (\ref{eq:falloff}) into account, the integrals
in (\ref{eq:omega}) can be numerically calculated. With the Mercury values
($\Delta \Phi _0 = 0.502 \times 10^{-6}$, $T_0 = 87.969 \, \mathrm{d}$, 
$r_1=46 \, 001 \, 200 \, \mathrm{km}$ and $r_2=69 \, 816 \, 900 \, \mathrm{km}$) we
find
\begin{equation}\label{eq:omegaphiA1}
\dfrac{\omega - \omega _0}{\omega _0} \, = \, 
3.3 \times 10^{-1}\, \phi _{11} \, + \, 
3.1 \times 10^{-1} \, \phi _{21} \, + \, 
2.5 \times 10^{-8} \, \phi _{01} \; .
\end{equation}
The observational fact that the left-hand side is bounded by $10^{-3}$, 
toghether with (\ref{eq:phi01b}), implies that
\begin{equation}\label{eq:omphi11phi21}
\big| \, \phi _{11} \, + \, 
9.4 \times 10^{-1} \phi _{21} \, \big| 
\, \lessapprox \, 3.0 \times 10^{-3} \; .
\end{equation}
In combination with (\ref{eq:phi11phi21b}) this gives us a bound for
$\phi _{21}$,
\begin{equation}\label{eq:boundphi21}
\big| \,  \phi _{21} \, \big| \, \lessapprox \, 3.6 \times 10^{-3} \; ,
\end{equation}
which means that the Finslerity is bounded by
\begin{equation}\label{eq:boundphi2r}
\big| \,  \phi _{2} (r) \, \big| \, \lessapprox \, 1.8 \times 10^{-10} \; 
\end{equation}
everywhere in the Solar system beyond the Mercury orbit. 

\section{Conclusions}\label{sec:conclusionss}
In this paper we have considered a class of spherically symmetric and static
Finsler spacetimes which are small perturbations of the Schwarzschild metric.
After fixing the ambiguity in the choice of the radial coordinate by requiring
that a sphere at coordinate $r$ has area $4 \pi r^2$, the perturbed metric 
is characterised by three functions $\phi _0(r)$, $\phi _1(r)$ and $\phi _2(r)$ which we
called the ``perturbation functions''. It was our main goal to determine the bounds 
which are imposed on these perturbation functions by observations in the Solar 
system. In this way we have provided a framework for testing if a certain Finsler 
modification of general relativity is in agreement with experimental facts. 

We have been careful to set up the formalism in such a way that not only 
freely falling particles but also light rays are unambiguously defined as Finsler 
geodesics. We feel that this is a major advantage in comparison to several 
other Finsler approaches where the definition of light rays is questionable.  
Having both freely falling particles and light rays at our disposal is essential
because these are the tools needed for the experiments discussed.

The formalism presented here is meant as an analogue of the PPN formalism. 
From a methodological point of view, there are two differences.
First, our formalism is post-Schwarzschild rather than post-Newtonian. This
is, of course, motivated by the fact that we wanted to concentrate on the
possible Finsler deviations from standard general relativity. Second, we
chose for the radial coordinate the area coordinate, whereas in the standard
PPN formalism one chooses the isotropic radial coordinate. This is a necessary
deviation from the standard PPN formalism because the isotropic radial coordinate
does not exist in a proper Finsler spacetime. 

Our approach is purely kinematical, 
i.e., no field equation is used. Hence, one can use it for testing the validity
of solutions to \emph{any} Finslerian field equation, provided that the solutions
belong to the class considered in this paper.

Here we have discussed only tests where all objects moving in the field
of the Sun can be treated as test particles. One could set up a Finsler
geometry model for more complicated situations, e.g. for the motion of
the Earth in the combined gravitational field of the Sun and the Moon. 
This would make more sensitive tests possible, in particular using the 
very precise Lunar Laser Ranging measurements. This is planned to be
done in future work. 
  
\section*{Appendix} 
For our work it was crucial that the timelike Finsler geodesics were
interpreted as freely falling particles and the lightlike geodesics 
were interpreted as light rays. We feel that this is the most natural
interpetation, if one believes in the possibility that our real world
carries a Finsler spacetime structure. As far as the light rays are
concerned, the geodesic hypothesis can be further justified by 
deriving the lightlike geodesics as the bicharacteristic curves of
appropriately generalised Maxwell equations. In this appendix
we will outline how this can be done. As a detailed treatment
would require a separate paper, we will only sketch the line of
thought, just enough to convince the reader that the geodesic
hypothesis for light rays is, indeed, well motivated.

In Definition \ref{def:Finsler} we have defined Finsler spacetimes in
terms of a Lagrangian $\mathcal{L}(x , \dot{x})$. We can switch to a Hamiltonian
formulation by introducing canonical momenta
\begin{equation}\label{eq:momenta}
p_{\mu} = \dfrac{\partial \mathcal{L}(x , \dot{x} )}{\partial \dot{x}{}^{\mu}}
= g_{\mu \nu} (x, \dot{x} ) \dot{x}{}^{\nu}
\end{equation}
and the Hamiltonian
\begin{equation}\label{eq:Hamilton}
\mathcal{H}(x,p) = p_{\mu} \dot{x}{}^{\mu} - \mathcal{L} (x, \dot{x} ) =
\dfrac{1}{2} \, g^{\mu \nu} (x, p ) p_{\mu} p_{\nu} 
\end{equation}
where $g^{\mu \nu} (x,p)$ is defined through
\begin{equation}\label{eq:gcontra}
g^{\mu \nu} (x,p) g_{\nu \sigma} (x , \dot{x} ) = 
\delta ^{\mu} _{\sigma} \; .
\end{equation}
In (\ref{eq:Hamilton}) and (\ref{eq:gcontra}), $\dot{x}{}^{\mu}$
must be expressed as a function of $x$ and $p$ with the help
of (\ref{eq:momenta}). For later convenience, we write
\begin{equation}\label{eq:Hmu}
\mathcal{H}^{\mu} (x,p) = \dfrac{\partial \mathcal{H}(x,p)}{\partial p_{\mu}} \; .
\end{equation}
Note that, because $\mathcal{L} ( x , \dot{x})$ is assumed to be homogenoeus
of degree two with respect to the $\dot{x} {}^{\mu}$, the Hamiltonian
$\mathcal{H}(x,p)$ is homogeneous of degree two with respect to the $p_{\mu}$,
hence
\begin{equation}\label{eq:HmuH}
p_{\mu} \mathcal{H}^{\mu} (x,p) = 2 \,  \mathcal{H}(x,p) \; .
\end{equation}
The lightlike Finsler geodesics are the solutions to Hamilton's equations
with $\mathcal{H}(x,p)=0$. It is our goal to demonstrate that these are
the bicharacteristic curves of appropriately generalised Maxwell equations.

In the case of a pseudo-Riemannian metric, where the $g^{\mu \nu}$
depend on $x$ only and not on $p$, the source-free vacuum Maxwell
equations can be written as
\begin{equation}\label{eq:Max1}
\partial _{\mu} F_{\nu \sigma} (x) +
\partial _{\nu} F_{\sigma \mu} (x) +  
\partial _{\sigma} F_{\mu \nu} (x) = 0
\end{equation}
\begin{equation}\label{eq:Max2}
\big( \mathcal{H}^{\mu} (x, \partial ) \, + \, \dots \, \big) F_{\mu \nu} = 0 \; .
\end{equation}
Here $F_{\mu \nu}$ is the electromagnetic field tensor and $\partial _{\mu}$
means partial derivative with respect to $x^{\mu}$. In (\ref{eq:Max2})
we have written only the principal part. $\mathcal{H}^{\mu} (x, \partial) = g^{\mu \sigma} (x)
\partial _{\sigma}$ is a first-order differential operator and, thus, homogeneous
of degree one with respect to the $\partial _{\sigma}$. The omitted terms, indicated
by ellipses, involve the Christoffel symbols and no derivatives, so they are in
particular homogeneous of degree zero with respect to the $\partial _{\mu}$.

It is very natural to postulate that Maxwell's equations take the same form of
(\ref{eq:Max1}) and (\ref{eq:Max2}) on a Finsler spacetime. Now $\mathcal{H}^{\mu} (x,p)$
is no longer a polynomial, but still homogeneous of degree one with respect to
the $p_{\mu}$, so $\mathcal{H}^{\mu} (x , \partial )$ is no longer a differential operator
but still a well-defined pseudo-differential operator. (For a detailed exposition of
pseudo-differential operators see, e.g., H{\"o}rmander \cite{Hormander1985}.)
The terms indicated by ellipses
are necessary in (\ref{eq:Max2}) to make this equation coordinate-independent.
Their special form, however, will not be relevant for the following argument. We
only have to assume that they are homogeneous of degree one with respect to
the $\partial _{\mu}$, as in the pseudo-Riemannian case.     
  
We now apply the operator $\mathcal{H}^{\mu}(x, \partial )$ to (\ref{eq:Max1}). With the
help of (\ref{eq:Max2}) and (\ref{eq:HmuH}) we find
\begin{equation}\label{eq:wave}
\mathcal{H}(x, \partial) F_{\nu \sigma} \, + \, \dots \, = \, 0 
\end{equation}
where again only the principal part (i.e., the terms of highest degree of homogeneity)
has been written out. This demonstrates that the field tensor satisfies a (pseudo-differential)
Finslerian wave equation. The principal part determines the \emph{characteristic equation}
(or \emph{eikonal equation})
\begin{equation}\label{eq:eikonal}
\mathcal{H}(x, \partial S ) = 0 \; .
\end{equation}
It gives the characteristic surfaces $S= \mathrm{const}$ along which solutions 
$F_{\mu \nu} (x)$ to Maxwell's equations  might have discontinuities of their
first derivatives. The characteristic equation has the form of a Hamilton-Jacobi 
equation with the Finsler Hamiltonian $\mathcal{H}(x,p)$. The \emph{bicharacteristic curves}
(or \emph{rays}) are the corresponding solutions to Hamilton's equations, i.e., 
the lightlike Finsler geodesics.

We have thus derived the result that light rays are lightlike Finsler geodesics from
the assumption that Maxwell's equations on a Finsler spacetime take the form of
(\ref{eq:Max1}) and (\ref{eq:Max2}). A full treatment of the subject would, of course,
require to specify the omitted terms in (\ref{eq:Max2}) and to discuss their
implications for physics. This could be the subject of another paper.
  
\section*{Acknowledgments} 
CL and VP gratefully acknowledge support from the Deutsche 
Forschungsgemeinschaft within the Research Training Group 1620
``Models of Gravity''. Moreover, VP was financially supported 
by the German-Israeli Foundation, Grant 1078/2009, and by
the Deutsche Forschungsgemeinschaft, Grant LA905/10-1, during
the course of this work. 
CL wishes to thank the cluster of excellence QUEST for support 
and J{\"u}rgen M{\"u}ller for discussions.

\end{document}